# An investigation on electronic and magnetic properties of Cr substituted MoS$_2$ monolayer and multilayers – Hybrid functional calculations

Aloka Ranjan Sahoo[1,2] and Sharat Chandra[1,2]

[1]Materials Science Group, Indira Gandhi Centre for Atomic Research, Kalpakkam, 603102, Tamil Nadu, India

[2]Homi Bhabha National Institute, Mumbai, 400094, India

Corresponding authors email: alokranjanctc@gmail.com, sharat@igcar.gov.in

## Abstract

With help of ab-initio density functional theory calculation, DFT+U, and hybrid functional HSE06, we revisit the layer-dependent electronic structure and magnetic properties of pristine and 3$d$ transition metal Cr doped MoS$_2$ monolayer and multi-layers. Our results show that the dopant Cr atoms prefer to stay at nearest neighbor distances. In the multilayers, they prefer to remain in the outermost surface layers. Matching with the experimental band gap, the optimized U parameter we report is 4 eV. The band gap of the Cr-doped monolayer is indirect, confirming the experimental observation from photoluminescence experiments. The HSE06 calculation for Cr doped monolayer shows that the band gap of doped Cr MoS$_2$ monolayer is indirect and no magnetism is observed. From the DFT studies, the band gap for the multilayers is indirect, and doping with Cr doesn't induce magnetic moments in MoS$_2$ layers. The band gap is observed to decrease with the multilayer thickness. The strain induced by substitutional Cr doping at the Mo site transforms the band gap in monolayer MoS$_2$ from direct to indirect; the defect states are produced within the band gap region close to the conduction band minimum.

# 1. Introduction

The successful micro-mechanical exfoliation of individual honeycomb atomic planes of carbon atoms in graphite leads to the discovery of a new two-dimensional (2D) material called grapheme [1, 2]. The 2D form of graphite, with its rich physics and fascinating properties [3–5], possesses high mechanical strength, carrier mobility, and a narrow band gap. This celebrated discovery of graphene has received much attention from researchers to explore other two-dimensional (2D) families of materials. Two-dimensional materials possess a sheet-like structure consisting of a few atoms thick that can be synthesized by experimental techniques such as mechanical exfoliation, liquid exfoliation, gas vapor growth, chemical synthesis, and other methods [1, 2, 6]. Due to the electronic confinement along one direction, 2D materials show different properties than their bulk counterparts [7, 8]. Other materials that have layered structures include hexagonal boron nitride (h-BN) [9], transition metal dichalcogenides (TMDs) [10–12], single-layer group-III mono-chalcogenides (MX), mono-elemental Xenes (phosphorene, arsenene, silicene, germanene, etc.) and ternary oxide. Materials from these families have received much attention from researchers to explore their properties in low-dimensional form. TMDs are often denoted by the chemical formula $MX_2$, where M represents the transition metal TM atom, and X denotes the chalcogen atom. The TM atom, M, is sandwiched between two triangular planes of chalcogen atoms X in a tetrahedral arrangement that forms one layer of $MX_2$, i.e., X-M-X. The chalcogen X and transition metal M in the X-M-X sandwiches have covalent bonding, whereas the layers of X-M-X sandwiches interact via weak Van der Waals force [13–14]. In the band structure of TMDs, contribution to the valence band maxima (VBM) and conduction band minimum (CBM) comes from the $d$-orbitals of transition metal (TM). Due to different electronic occupancy of $d$-orbitals in the TM atoms, different TM-based TMDs show diverse electronic properties [15]. Depending on the *d*-orbital electron occupancy of TM, TMDs can be metal, semiconductor, insulator, superconductor, or topological insulator [16]. Among this family, the group-VI transition metal dichalcogenides, $WS_2$, $VS_2$, $MoSe_2$, $MoS_2$, $VSe_2$, etc., are the extensively studied materials for their application in electronic, optoelectronic, and photovoltaic devices [17] for next-generation nanoelectronics and nanophotonics applications [2, 7, 18–21].

Molybdenum disulfide ($MoS_2$), an essential member of the TMD family, can exist in the 1T, 2H, and 3R phases. Its 2H phase has a space group of $P6_3mmc$ and point group $D_{6h}$, and it is semiconducting with an indirect band gap of 1.2 eV [22, 23]. In $MoS_2$, hexagonally arranged Mo atoms are sandwiched between two hexagonal planes of S atoms in the trigonal prismatic arrangement. The unit of this arrangement forms the $MoS_2$ monolayer. The electronic properties of 2D $MoS_2$ show a strong dependence on layer thickness. The band gap in multilayer $MoS_2$ is indirect, and the indirect band gap widens upon decreasing the number of layers. In the monolayer limit, the band gap is 1.9 eV [7, 20, 24, 25] which is direct. This transition from an indirect band gap in bulk to a direct band gap in monolayer is attributed to the absence of inversion symmetry in monolayer [7]. Different experimental methods, such as micromechanical cleavage, mechanical exfoliation, liquid exfoliation, and chemical vapor deposition (CVD), have been used to synthesize 2D $MoS_2$ [2, 18, 26, 27].

Although room temperature quantum hall effect and transport are observed in grapheme [4] experimentally, the absence of a band gap causes hindrance for switching action in the circuit and control of electron flow. Although these difficulties can be overcome by modifying the substrate while depositing [28] or making nano-ribbons [29, 30], it deteriorates the carrier mobility [29, 31]. In contrast to this, the monolayer $MoS_2$ offers quite distinct properties like a high current on-off switching rate ($10^8$), high carrier mobility (~200 cm$^2$/Vs), high thermal stability in the field effect transistor [21], tunable band gap, optical and photocatalytic properties [32], catalytic properties [33], etc.

MoS$_2$ has a wide range of applications, such as in sensors [34], dry lubricants [35], catalysts [36], photovoltaic [37], energy storage [38] and spintronics device applications [39–42]. Theoretical calculations have predicted the spin Hall effect and spin manipulation by the electric field in the MoS$_2$ monolayers [43–45]. Thus, the 2D molybdenum dichalcogenide (MoS$_2$) is a promising candidate material for nano-electronics and spintronics applications due to its ultra-thin surface [46], easy fabrication method, surface free of dangling bonds, charge transport properties [47], and tunable electronic properties with layer-thickness [7]. Further, the properties of 2D MoS$_2$ can be tuned by external means as well, viz., electric field, defects, strain [48], etc. Specifically, magnetism can be induced in the monolayer MoS$_2$ by surface functionalization, applying strain [49–51], and creating vacancy defects [51]. In general, defects play a significant role in altering the physical and chemical properties of materials. It can help to change electronic properties like carrier mobility, induce magnetism, and improve the catalytic properties of the host material [33]. It has been experimentally reported that the Mn-doped MoS$_2$ nano-flakes [52] show long-range ferromagnetic ordering for the relatively high doping concentration [53], supported by simulations. The calculations based on Monte Carlo simulation predicted that the Curie temperature of Mn-doped at the Mo site in monolayer MoS$_2$ can be above room temperature when doping concentration is 10–15% [54]. The other TM atoms (Fe, Co, Ni, Cr, etc.) are also employed for studying the magnetism in monolayer MoS$_2$ [55–57] and found that these magnetic elements can induce magnetic moments in monolayer MoS$_2$. Low-energy ion implantation can dope Mn, Fe, Co, and Ni into MoS$_2$ [58]. Fang *et al* discussed the various TM atom doping in

TMD families of material by CVD method [59], Lewis *et al* have doped Cr on MoS$_2$ thin films with aerosol-assisted CVD method [60], Zhang *et al* have found magnetism in Cr-doped MoS$_2$ nanosheets [61]. Apart from TM atoms, the adsorption of non-metal (NM) atoms like H, B, C, N, and F on pristine 1H-MoS$_2$ can also induce the local magnetic moment [62]. The dilute magnetic semiconductors (DMS) are essential in spintronics applications as they can be used for spin manipulation. Magnetism-induced MoS$_2$ behaves as the DMS. Cheng *et al* have reported the binding energies for TMDs and found it is similar to Mo for TM of groups 4, 5, and 6 [57].

In the literature, defect formation energy for various TM atoms doped in MoS$_2$ possible TM doping strategies for substitutional doping in monolayer MoS$_2$ [40, 55-57], have been discussed. Binding energies of IVB, VB and VIB group TM metal are similar [57] and increases with period. The magnetic properties of TM-doped monolayer MoS$_2$ have been studied using DFT and DFT+U methods [54, 55]. Cheng *et al* [57] have studied using the DFT+U method. Huang *et al* [63] successfully doped monolayer MoS$_2$ with Cr and Mn by CVD method and found that the doped MoS$_2$ has semiconducting nature. However, they have not discussed its magnetic properties. The reported results for magnetism in TM doped MoS$_2$ with DFT+U calculations are yet to be validated with experimental findings like the band gap value or magnetic moments. There is not sufficient focus on the substitutional doping of multiple TM atoms in monolayers and multilayers and the energetically stable configurations for the doped atoms.

Here we report layer dependent electronic structure of two Cr atoms doped in MoS$_2$ monolayer and multilayers (up to 4 layer thick) with in-plane and out-of-plane doping. The optimal Cr-Cr distance in the in-plane monolayer, both the in-plane and out-of-plane doping in the multilayers and the preferred layer, i.e whether the dopant atom likes to stay on the outer or inside layers in the multilayers, defect levels, band edges and possible magnetism in the Cr-doped monolayer have been studied by both DFT+U and HSE06 hybrid functional calculations. The direct band gap in MoS$_2$ monolayer and its change to the indirect band gap in Cr-doped MoS$_2$ monolayer have been observed in experiments.

## 2. Calculations details

The structural optimization and electronic structure calculations are performed for monolayers and multilayers of MoS$_2$ using the first principles method, the density functional theory, as implemented in the VASP [64, 65]. The defect supercells are obtained from the host supercell by substitutionally doping the Cr atom on Mo sites. Thus the total number of the atoms in the supercell remains unchanged and the total energies of the host and defect supercells can be compared directly. The convergence studies for finding appropriate supercell size are carried out to minimise the image-image interaction between the dopant Cr atoms. Supercell sizes ranging from 1×1×1 to 6×6×1 are considered. The difference between total energy per atom of the defect supecell and the host supercell of same size is computed and compared for different supercell sizes. It is seen that the total energy per atom varies from 0.38 eV/atom for the 1×1×1 supercell to 0.02 eV/atom 6×6×1 supercell, with the 4×4×1 supercell having the energy of 0.05 eV/atom. We have chosen the 4×4×1 supercell for all the calculations. The supercell of monolayer contains 48 atoms (16 Mo and 32 S) in total, while the multilayers contain 96 atoms for bilayers, 144 atoms for trilayers, and 192 atoms for four layers. For doping Cr atoms in the layers, there are only three unique in-plane configurations with two Cr atoms in the same plane and four unique out-of-plane configurations where two Cr atoms are in two different planes. To sample the Brillouin zone, a Γ-point centered K-point mesh of 4×4×1 is used [66]. The plane wave energy cutoff was set to 450 eV. PAW pseudopotentials with GGA exchange-correlation functional describe ion-electron interactions [67–69]. The energy convergence criterion for electronic self-consistency was set at $10^{-6}$ eV. A vacuum of 12 Å is taken along the $z$-direction to avoid unrealistic interactions between the periodic images of slabs. We have ensured that the total energy of the system doesn't change as a function of the vacuum size. For the multilayer systems, to account for the weak Van der Walls interaction between atomic layers, the Grimme D3 correction is used [70]. The optimized hexagonal lattice parameter of bulk $MoS_2$ is $a = 3.16$ Å and $c = 12.95$ Å, which is in excellent agreement with experimental lattice parameters. All the structures are relaxed until the forces on each ion are less than 0.01 eV/Å. Semi-empirical DFT+U [71, 72] studies are carried out to include the on-site Coulomb interactions for localized $3d$ and $4d$ electrons of Cr and Mo, with different U parameters viz., 0.0, 1.0, 1.5, 2.0, 2.7 3.0, 3.3, 3.5, 3.7 and 4.0 eV. HSE06 calculation with an exact exchange of 15% is included in the exchange functional [73] to match with experimental observations [63].

## 3. Results and discussion

### 3.1 Electronic structure of Bulk $MoS_2$ and multilayers

The calculated inter-layer spacing of hexagonal $MoS_2$ in the 2H phase is 6.16 Å. The in-plane Mo-Mo and Mo-S bond lengths are 3.16 Å and 2.40 Å, respectively. The calculated band structure and electronic density of states of bulk $MoS_2$ (2H phase) are plotted in Figure 1. The band structure shows that the valence band maximum (VBM) is at the $\Gamma$ point; however, the conduction band minimum (CBM) is in the K-$\Gamma$ direction, which means it is an indirect band gap. Contributions to the valence and conduction band edges are formed by strongly hybridizing the $4d$-orbital of Mo and $3p$-orbitals of S. The calculated band gap is 0.89 eV, close to the experimental band gap value of 1.2 eV and other reported calculations [22, 23]. The band gap is underestimated by 2.5% [71].

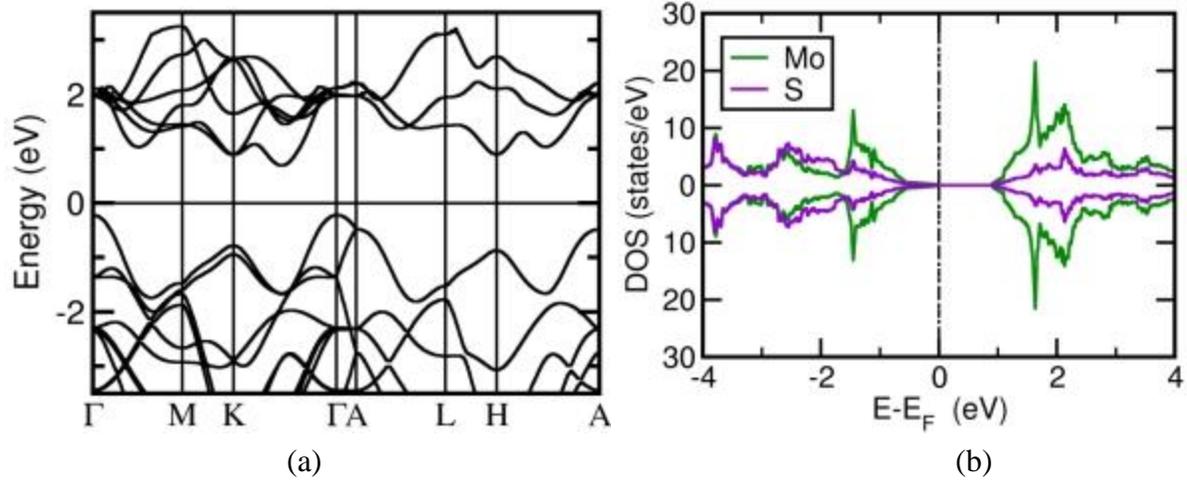

**Figure 1**. The (a) band structure and (b) atom projected density of states of bulk 2H-MoS$_2$

Thus, the calculations are performed using the hybrid functional to describe the electronic structure correctly. Figure 2 shows the band structure of bulk MoS$_2$ obtained with DFT and HSE06 calculation. For DFT calculations, the full bands are 24, of which 8 are empty conduction bands. At the same time, the HSE06 calculations are done by adding more empty conduction bands, totaling the number of bands to 40. In the HSE06 functional-based calculation, the fraction of exact exchange used is 15%. This results in the shifting of the conduction band by 0.34 eV to higher energy. The band gap of 1.24 eV matches the experimental band gap of 1.24 eV.

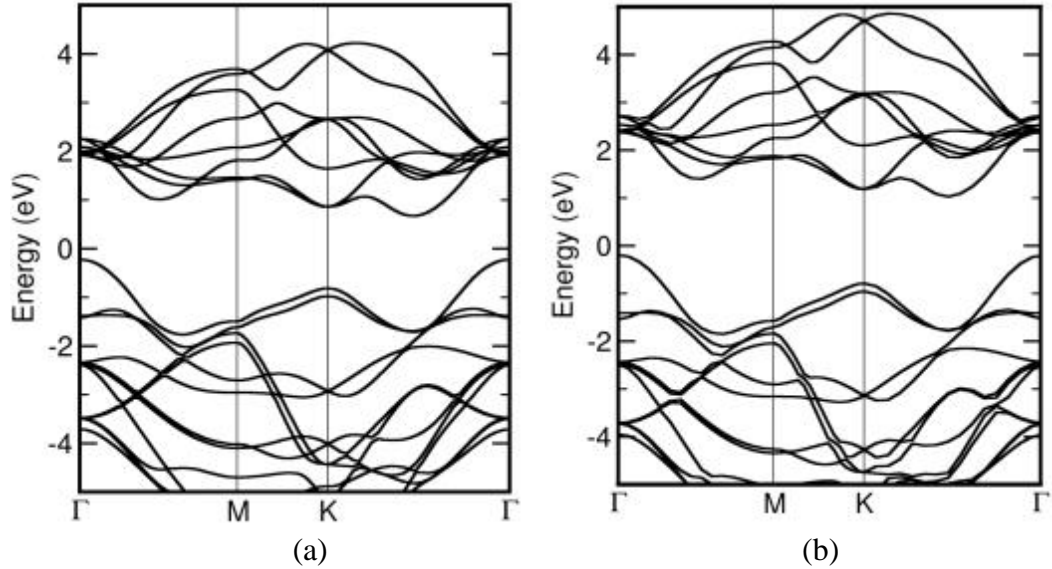

(a)                  (b)

**Figure 2**. Band structure of bulk $MoS_2$ with (a) DFT and (b) HSE method.

### 3.2 Undoped $MoS_2$ Multilayers

Next, we have compared the electronic structures of the different numbers of layers (one to four) of $MoS_2$. The atom-projected DOS and band structure for monolayer, bilayer, trilayer, and four-layers of $MoS_2$ are shown in Figure 3 and Figure 4, respectively. In contrast to bulk, the monolayer has a direct band gap, where both VBM and CBM are at high symmetry points, K. The transition in the monolayer is due to the absence of inversion symmetry in the single layer. When the number of layers is two or more in multilayers, the band gap is indirect and decreases with an increase in the number of layers. This is because the band contribution at $\Gamma$ point comes from $d_{z^2}$ orbitals. In contrast, at K point, the contribution to the band comes from $d_{x^2-y^2}$ and $d_{xy}$ orbitals, therefore in the multilayer configurations, the inter-planar van der Walls interactions come into play and shift the band at the $\Gamma$ point [16]. The band gap for the mono-layer is 1.74 eV, close to the experimental value of 1.9 eV [7]. The band gap values are tabulated in Table 1 for the monolayers and multilayers.

**Table 1.** The energy band gap of various 2D and bulk MoS$_2$. The monolayer has a direct band gap, whereas it is indirect for the bulk and multilayers.

| System | Monolayer | Bilayer | Trilayer | Four-layer | Bulk |
|---|---|---|---|---|---|
| Band Gap | 1.74 eV (Direct) | 1.30 eV (Indirect) | 1.10eV (Indirect) | 1.03 eV (Indirect) | 0.89 eV (Indirect) |

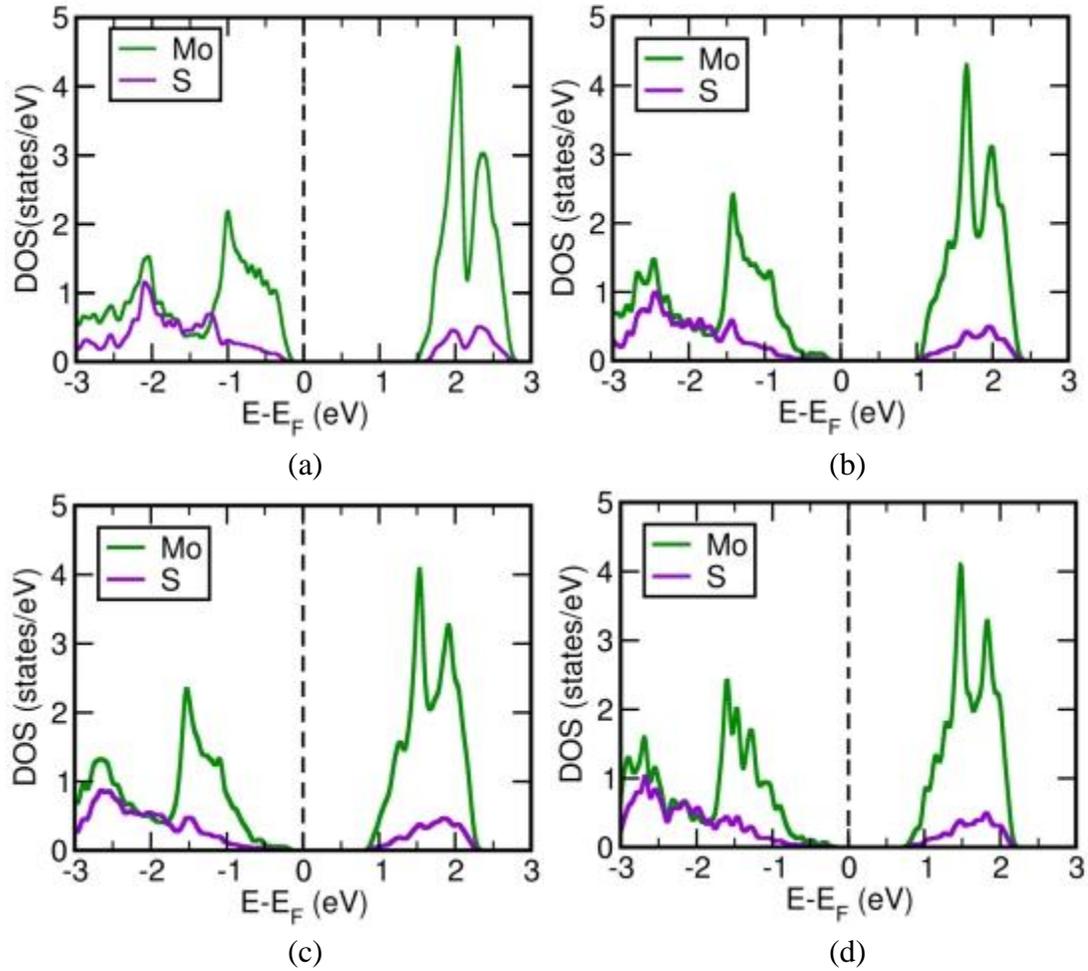

**Figure 3.** Atom projected DOS for undoped MoS$_2$ (a) monolayer, (b) bilayer, (c) trilayer, (d) four-layer.

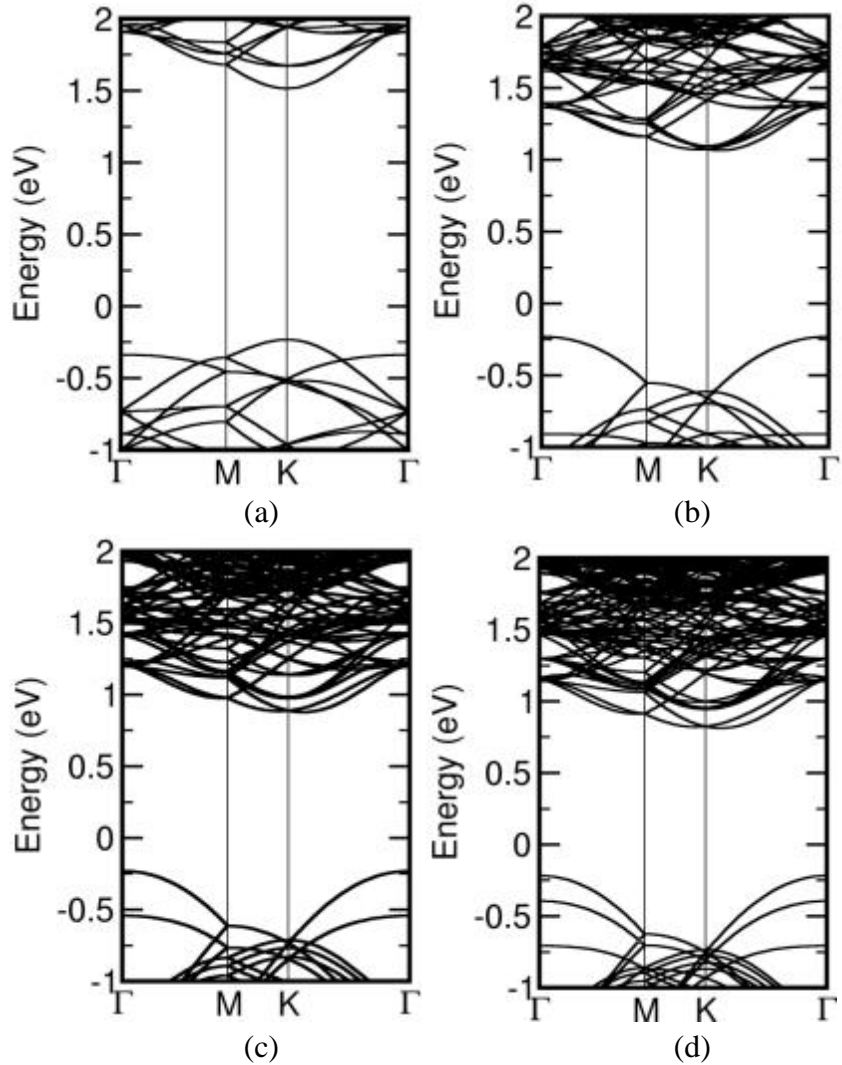

**Figure 4**. Band structure of mono-layer (a), bilayer (b), trilayer (c), and four-layer (d) MoS$_2$.

### 3.3 Different configurations for dopant sites

Further, we dope the Cr in the bulk and 2D systems of MoS$_2$ at the site of Mo. The two Cr atoms are doped in two ways for multilayers, with three in-plane and four out-of-plane configurations. Cr atoms are doped in the in-plane configuration for the bulk and monolayer.

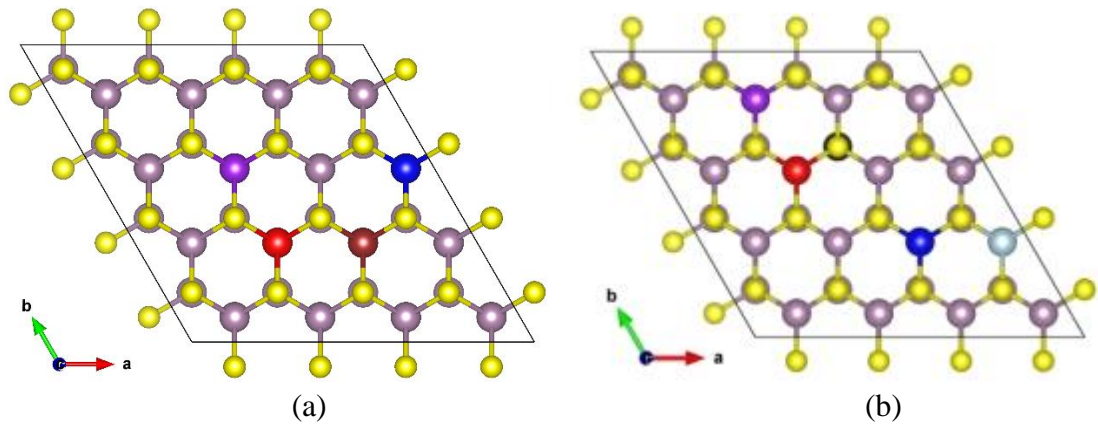

(a)                               (b)

**Figure 5**. The unique possible positions for two Cr dopants for (a) in-plane (b) out-of-plane configurations in a 4×4×1 supercell of $MoS_2$. The in-plane case has three configurations, whereas the out-plane case has four unique configurations. The Mo atoms are in violet and S in yellow, while other colors are described in the text.

The possible unique in-plane configurations for a 4×4×1 supercell of $MoS_2$ are shown in Figure 5. The possible unique configurations for in-plane doping for the first dopant at the site denoted by the purple ball, and the possible unique sites for the second dopant are shown by the red, blue, and brown ball (Figure 5a). For the first dopant at the site represented as black ball, the number of unique sites possible for the second dopant in out-of-plane configuration is shown with balls with different colors, viz, red, blue, cyan and purple (Figure 5b).

**Table 2**. The total energy for various configurations in the in-plane and out-of-plane doping.

| Configurations | 1 | 2 | 3 | 4 | 5 |
|---|---|---|---|---|---|
| In-plane | -716.8083 eV | -716.7538 eV | -716.7734 eV | -716.8031 eV | -- |
| Out-plane | -716.8104 eV | -716.8102 eV | -716.7974 eV | -716.7818 eV | -716.7742 eV |

The total energies of the configurations are given in Table 2. The lowest energy configurations correspond to the cases where the Cr atoms are in the nearest neighbor positions. For the in-plane configurations, the optimal Cr-Cr distance is 3.14 Å, and for the out-of-plane configurations, it is

6.41 Å. Only the results of the lowest energy configurations (1 in both cases) are discussed in this paper.

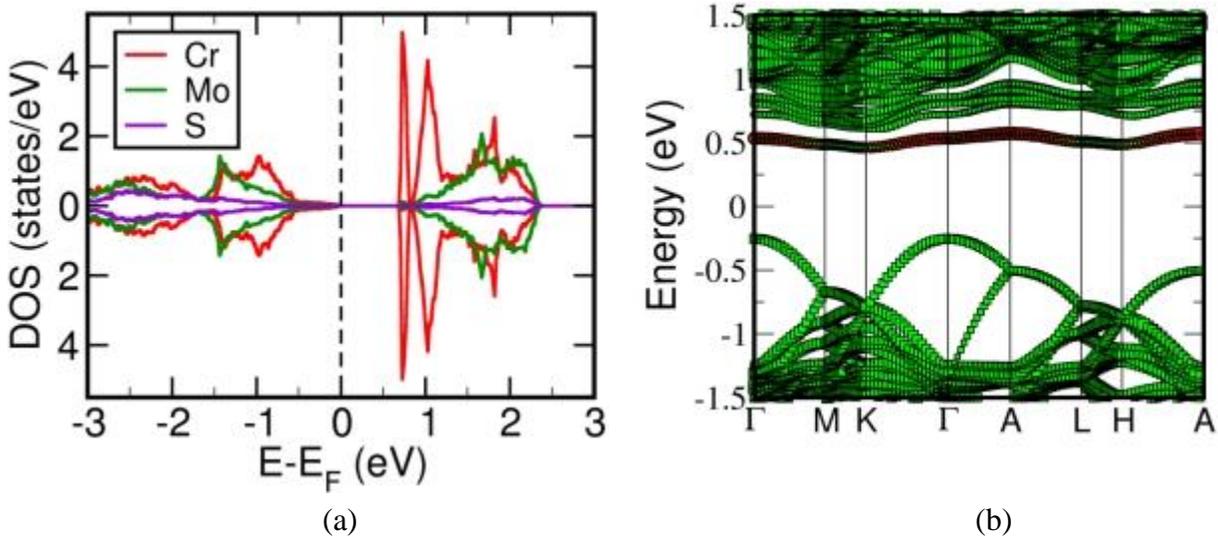

(a) (b)

**Figure 6.** The electronic density of states (a) and band structure for Cr-doped bulk MoS$_2$ (b).

Figure 6 shows the electronic structure for Cr doping in the in-plane configuration of bulk MoS$_2$. The bulk MoS$_2$ supercell (4×4×1) is doped with two Cr, corresponding to 6.25% doping. The Cr doping changes the band structure of the host and a Cr defect level is formed at 0.5 eV above the Fermi level, which is 0.3 eV below the conduction band of pristine MoS$_2$. The Cr-doped bulk MoS$_2$ is semiconducting and remains indirect with a band gap value of 0.85 eV. The band structure for MoS$_2$ monolayers with and without Cr doping is shown in Figure 7. The Cr doping leads to new levels in the band structure of the MoS$_2$ monolayer between 1.0 eV to 1.75 eV above the Fermi level. The defect level at ~1.2 eV above the Fermi level is due to Cr $d$-orbitals, whereas bands above this have equal contributions from Cr and Mo.

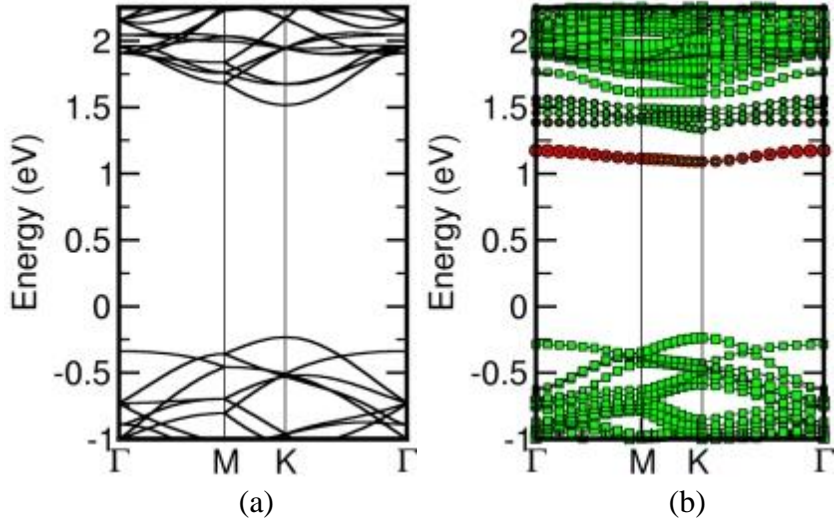

**Figure 7**. The band structure for a monolayer of $MoS_2$ (a) pristine and (b) two Cr atoms substituted at Mo sites. In the doped case, the contribution due to Cr is shown by bands with red colour and green for Mo atoms.

Like the bulk and monolayer, the atom projected DOS for the in-plane and out-of-plane configurations in the multilayers are shown in Figure 8 for the Cr-doped bi, tri, and four layers of $MoS_2$. The band structures for Cr-doped in-plane configurations of the monolayer, bilayer, trilayer, and four-layer are shown in Figure 9, and Cr-doped in out-of-plane configurations are shown in Figure 10. The Cr defect levels are formed at 0.65 eV, 0.6 eV, and 0.5 eV above the Fermi level for the in-plane doping in the bi, tri, and four-layer configurations. These defect levels are formed by the Cr $d$-orbitals. The splitting of energy levels in the conduction band increases with an increase in layer-thickness, leading to larger bandwidth of the Cr defect level. The indirect band gaps for the multilayers are 1.92 eV (monolayer), 1.32 eV (bilayer), 1.10 eV (trilayer), and 1.02 eV (four-layer), respectively.

There is no change to valance band energy levels. Cr-defect levels are at 0.1 to 0.5 eV below the conduction band for the Cr-doped monolayer. The band gap of in-plane Cr-doped multilayers is indirect where the VBM is at Γ, and the CBM is at K high symmetry point. For the Cr-doped $MoS_2$ monolayer, both the VBM and CBM are at high symmetry K-point. The TM-$d_{z^2}$ and S-$p$ orbitals contribute to the bands at Γ points, whereas the TM-$d_{xy}$ orbitals contribute to the bands at K-point. As the layer-thickness decreases, the interlayer interaction vanishes in the monolayer, resulting in both the VBM and CBM occurring at the K-point.

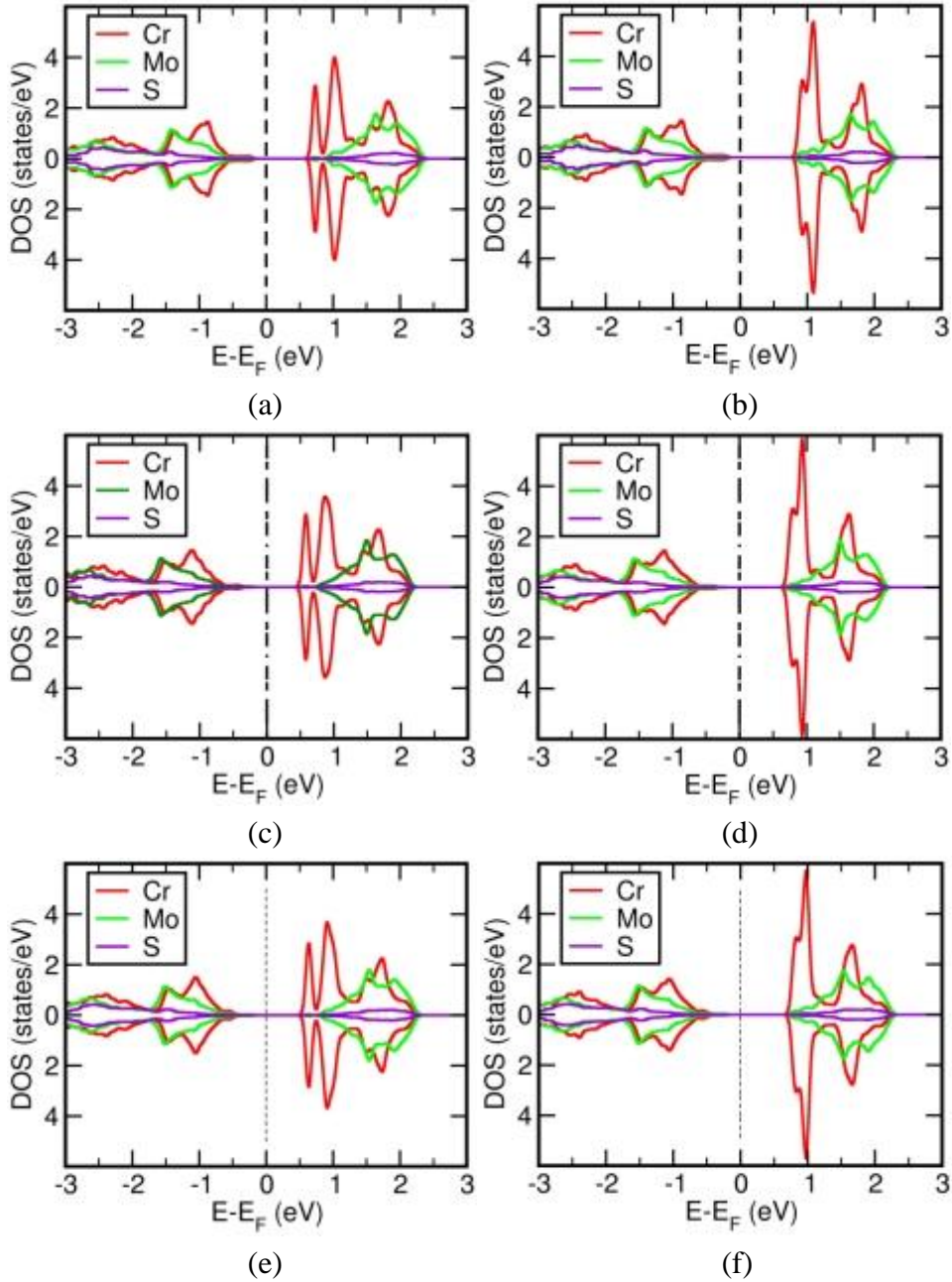

**Figure 8**. The DOS of the Cr-doped in-plane (a) bilayer and (b) trilayer and (c) four-layer MoS$_2$ and out-of-plane MoS$_2$ (d) bilayer (e) trilayer and (f) four-layer configurations, respectively.

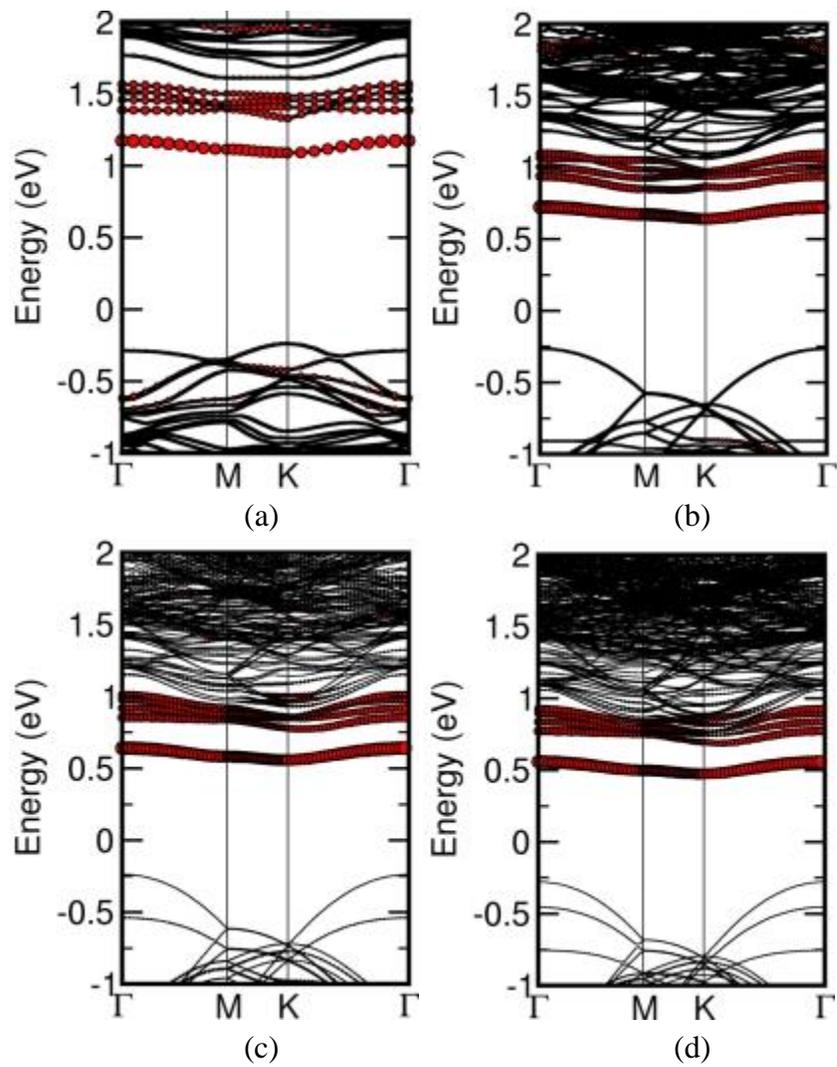

**Figure 9**. Band structure of Cr-doped in the in-plane of $MoS_2$ monolayer and multilayers, (a) monolayer, (b) bilayer, (c) trilayer, (d) four-layer.

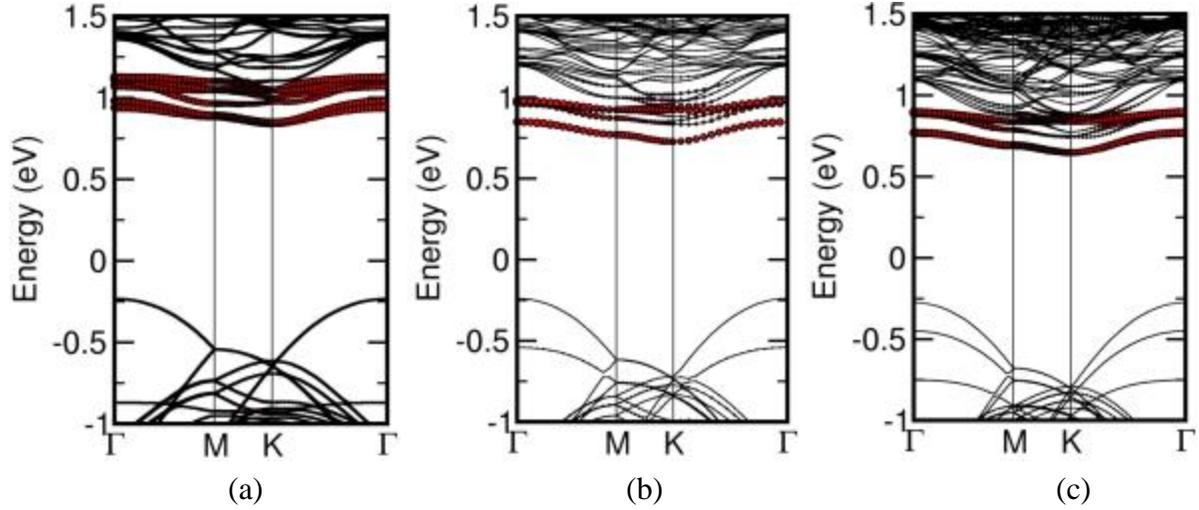

**Figure 10**. Band structure of Cr-doped out-of-plane configurations of $MoS_2$ multilayers, (a) for bilayer, (b) trilayer and (c) four-layer.

Cr-defect levels form at 0.2 eV below the conduction band for bilayer, trilayer, and four-layer out-plane Cr-doped $MoS_2$ multilayers. These defect levels are from Cr-$d$ orbitals. The band gap is indirect with the VBM at Γ-point and CBM at the K-point of high symmetry. Compared to the out-of-plane doping, the defect levels in the in-plane doped case are formed much below the conduction band. In the in-plane case, the Cr-$d$ orbital interacts strongly, whereas, in the out-of-plane case, the doped Cr atoms interact via the Cr $d_{z^2}$-orbitals only. As the doped Cr atoms for out-plane configuration stay on the surface of $MoS_2$, the interaction between them is weak, leading to deeper defect levels for the in-plane Cr-doped $MoS_2$ layers. Cr-defect levels are formed above 1.1 eV for the in-plane Cr-doped monolayer, and for the out-of-plane Cr-doped monolayer, the Cr defect levels are formed around 1.0 eV. In both cases, the contribution to the valance and conduction bands comes from Mo only, as seen in Figure 11.

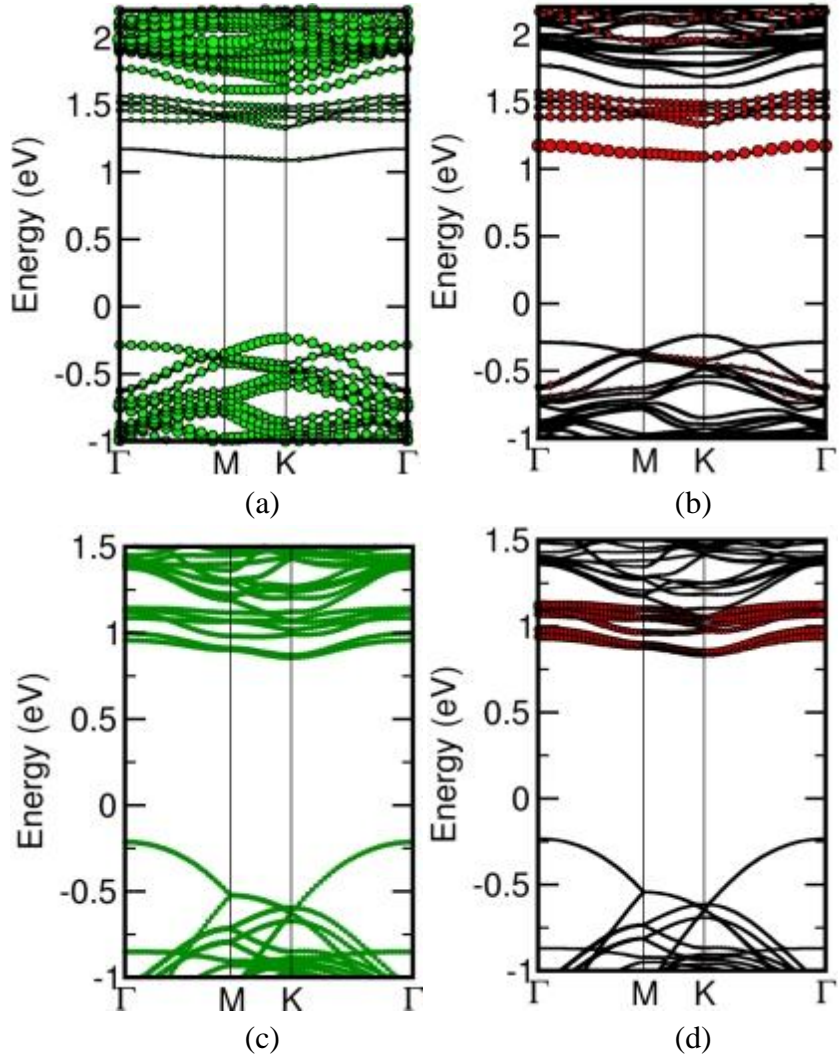

**Figure 11**. Atom projected band structure of in-plane Cr-doped MoS$_2$ monolayers (a) and (b), and out-of-plane Cr-doped bilayer MoS$_2$ (c) and (d). Green lines denote energy levels for Mo, and (b) red lines for energy levels for the Cr atom.

### 3.4 DFT+U study on monolayer MoS$_2$

Andriotis *et al* [54] studied the effects of the U parameter in doped MoS$_2$ monolayers using U = 0, 2.5, and 5.5 eV, suggesting the tunability of band gap in doped MoS$_2$ systems. However, the experimental band gap data were not available then, so there was uncertainty in the value of the U parameter. Wu Maokun *et al* [74] also did similar studies on 3$d$ TM doped MoS$_2$ monolayer and found that with U = 4.0 eV, the dope Cr atoms have a magnetic moment of 2.37 μ$_B$ and are antiferromagnetically coupled with each other. Although magnetism is observed with different U parameters, the band gap obtained is less than the reported experimental band gap for the undoped monolayer. By adding on-site Coulomb interaction to highly localized *3d* orbitals of Cr in Cr-doped MoS$_2$ monolayers, we observed that although Cr is a magnetic impurity, Cr doped monolayer in dilute concentrations exhibits no magnetic moment for U ≤ 3.5 eV. The values of the magnetic moment on Cr, the total magnetic moment, the band gap, and whether it is direct or indirect in nature are shown in Table 3. The magnetic coupling between the nearest neighboring Cr atom is antiferromagnetic for larger U, confirming earlier reports [54, 74]. Local distortion around the Cr atom leads to the redistribution of the energy levels of the octahedral bonded system (crystal field spin splitting). For U parameters of 3.7 eV and higher, the VBM is at high symmetry point K shifts to M and CBM at high symmetry direction of K, leading to the indirect nature band gap. This change can be attributed to the strain introduced by the dopant in the monolayer, as demonstrated by Conley *et al* [75] from photoluminescence studies, leading to the transition from the direct to indirect nature of the band gap. A magnetic moment of 0.2 $\mu_B$ on the six nearest neighboring Mo and a magnetic moment of 0.1 $\mu_B$ on the nearest S atoms are observed with antiferromagnetic ordering.

**Table 3**. Magnetic moments and band gaps obtained for different U parameter values in DFT+U calculations for Cr-doped MoS$_2$ monolayer with the Cr *3p*-orbitals in the core.

| U parameter, Cr (eV) | Magnetic moment, Cr ($\mu_B$) | Magnetic moment, total ($\mu_B$) | Band gap (eV) | Type |
|---|---|---|---|---|
| 0 | 0 | 0 | 1.92 | D |
| 1 | 0 | 0 | 1.93 | D |
| 1.5 | 0 | 0 | 1.93 | D |
| 2.0 | 0 | 0 | 1.92 | D |
| 2.5 | 0 | 0 | 1.92 | D |
| 3.0 | 0 | 0 | 1.93 | D |

| 3.3 | 0.1 | 0 | 1.92 | D |
| 3.5 | 0.0 | 0 | 1.92 | D |
| 3.7 | 2.5 | 0 | 1.76 | I |
| 4 | 2.62 | 0 | 1.75 | I |

In the DFT+U calculations, the 4$p$-orbitals of Mo are taken in valence band. If we also treat the 3$p$-electrons of Cr as the valence electrons, a significantly reduced magnetic moment of 0.48 $\mu_B$ on the Cr atom is observed. There is a magnetic moment of 0.12 $\mu_B$ on the four nearest neighboring Mo atoms in anti-ferromagnetic ordering. The material's band gap remains indirect, with a value of 1.88 eV. There is not much change in the other calculated properties.

### 3.5 HSE06 study on Monolayer $MoS_2$

The band gap for the undoped monolayer from the DFT calculation is 1.74 eV. We carry out the HSE06 hybrid functional calculations as the DFT with GGA couldn't give the correct band gap and magnetic ordering for the Cr-doped monolayer. We add 15% of the exact exchange to the exchange-correlation functional for the HSE06 calculations, which also need a higher number of conduction bands to describe the excited states correctly. Figure 12 shows the band structure for monolayer $MoS_2$ from DFT and HSE06 calculation. The HSE06 calculation gives a band gap of 1.90 eV for the undoped monolayer, where the conduction bands shift to higher energy by 0.25 eV. The VBM and CBM at high symmetry K point in both calculations show the direct nature of the band gap in the monolayer.

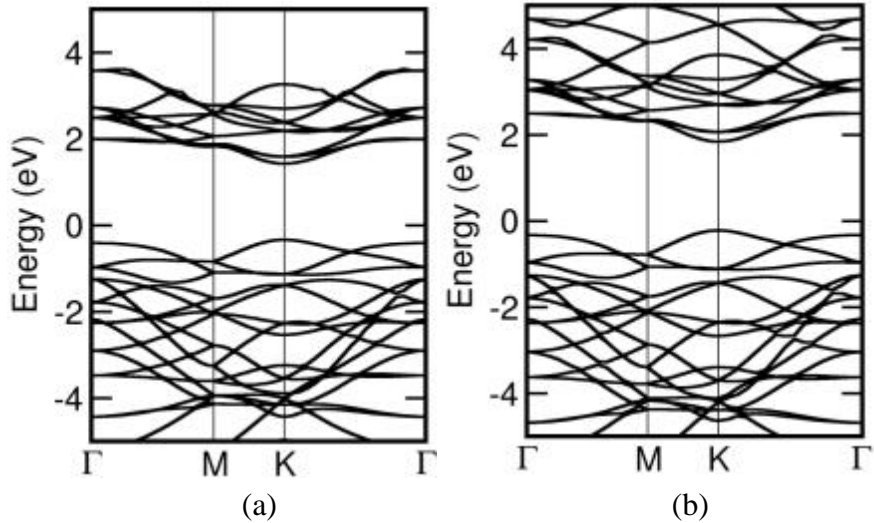

**Figure 12**. Band structure of monolayer MoS$_2$ with (a) DFT and (b) HSE06 hybrid functional calculation. With HSE06 hybrid functional, the conduction band shifts by 0.25 eV, and the energy levels are correctly described, which can be seen with the splitting of conduction band energy levels. The VBM and CBM are at the K-point, so the band gap is direct.

### 3.6 HSE06 for Cr doped MoS$_2$ monolayer

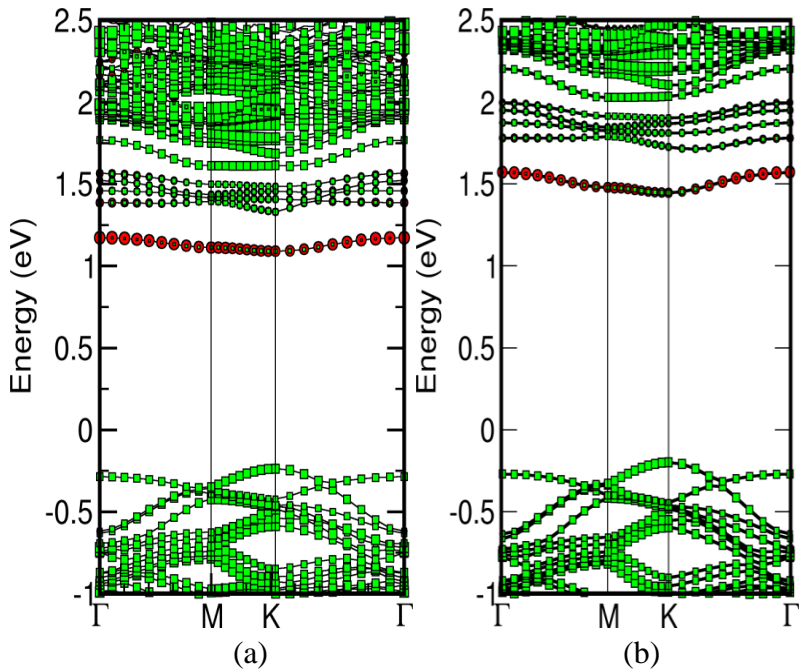

**Figure 13**. Band structure of Cr-doped monolayer MoS$_2$ with (a) DFT and (b) HSE06 calculation. With HSE06 hybrid functional, the Cr-defect level shifts upward and forms around 1.5 eV. The VBM is at the K point, whereas the CBM is at the K point in (a) and in between the K and Γ points in (b).

Figure 13 shows the band structure of Cr-doped monolayer MoS$_2$ obtained from the DFT and HSE06 calculations. For doped Cr MoS$_2$ monolayer, DFT estimates the band gap to be 1.92 eV for 12.5% doping. The Cr defect level is formed ~0.25 eV below the conduction band. The defect level gets contribution from the $3d_{z^2}$ orbital of the Cr atom, and other bands hybridize with Mo 4$d$ states. HSE06 functional with the exact exchange of 15% estimates the band gap to be 1.92 eV in agreement with the experimental photoluminescence measurements [63]. For both GGA and HSE06 functional the band gap is direct with both the VBM and CBM at the K point. The Cr defect levels contributed by $3d_{z^2}$ are formed ~0.25 eV below the conduction band.

## 4. Conclusions

In summary, energetically lowest energy configurations for in-plane doping of Cr in monolayer MoS$_2$ and in-plane as well as out-of-plane doping of Cr atom in multilayers of MoS$_2$ have been specified. Their electronic structures, DOS, and magnetic properties are calculated. The optimized configurations show that the doped Cr atoms prefer to occupy the nearest neighbor sites for the in-plane and out-of-plane doping. The doped Cr atoms prefer to stay in the surface layers for the in-plane doping. The Cr-Cr distance for in-plane doping is 3.14 Å and 6.14 Å for out-of-plane doping. Cr defects levels form below the conduction band within the band gap region, and the configurations are semiconducting. Both GGA and HSE06 predict that Cr-doped MoS$_2$ monolayers are semiconducting and non-magnetic with indirect band gaps. By including the on-site Coulomb interaction for the Cr *3d*-orbitals, we can induce magnetism in the system. The indirect nature of the band gap and exact band alignments for the in-plane Cr-doped monolayer MoS$_2$ is found from the calculations, which are supported by experimental findings.

## Acknowledgments

The authors would like to thank Dr. Gurpreet Kaur and Dr. Kashinath T. Chavan for discussions and technical help, and ARS would like to thank the IGCAR for the research fellowship grant.

## References

[1] Novoselov KS, Geim AK, Morozov SV, Jiang D, Zhang Y, Dubonos SV, *et al* 2004 Electric field effect in atomically thin carbon films. *Science* 306: 666-669

[2] Novoselov KS, Jiang D, Schedin F, Booth TJ, Khotkevich VV, Morozov SV, *et al* 2005 Two-dimensional atomic crystals. *Proc. Natl. Acad. Sci.* 102: 10451-10453

[3] Castro Neto AH, Guinea F, Peres NMRR, Novoselov KS, and Geim AK 2009 The electronic properties of graphene. *Rev. Mod. Phys*. 81: 109–162

[4] Zhang Y, Tan YW, Stormer HL and Kim P 2005 Experimental observation of the quantum Hall effect and Berry's phase in graphene. *Nature* 438: 201–204

[5] Novoselov KS, Geim AK, Morozov SV, Jiang D, Katsnelson MI, Grigorieva IV, *et al* 2005 Two-dimensional gas of massless Dirac fermions in graphene. *Nature* 438: 197-200

[6] Jin C, Lin F, Suenaga K, and Iijima S 2009 Fabrication of a Freestanding Boron Nitride Single Layer and Its Defect Assignments. *Phys. Rev. Lett.* 102: 195505

[7] Splendiani A, Sun L, Zhang Y, Li T, Kim J, Chim CY, *et al* 2010 Emerging photoluminescence in monolayer $MoS_2$. *Nano Lett*. 10: 1271-1275

[8] Li T, and Galli G 2007 Electronic Properties of $MoS_2$ Nanoparticles. *J. Phys. Chem. C* 111: 16192–16196

[9] Yang Z and Ni J 2010 Modulation of electronic properties of hexagonal boron nitride

bilayers by an electric field: A first principles study. *J. Appl. Phys.* 107: 104304

[10] Samadi M, Sarikhani N, Zirak M, Zhang H, Zhang HL and Moshfegh AZ 2018 Group 6 transition metal dichalcogenide nanomaterials: Synthesis, applications, and future perspectives. *Nanoscale Horizons* 3: 90-204

[11] Li X and Zhu H 2015 Two-dimensional MoS$_2$: Properties, preparation, and applications. *J. Materiomics* 1: 33–44

[12] Zhao X, Dai X, Xia C, Wang T and Peng Y 2015 Electronic and magnetic properties of Mn-doped monolayer WS$_2$. *Solid State Commun.* 215-216: 1-4

[13] Mattheiss LF 1973 Band structures of transition-metal-dichalcogenide layer compounds. *Phys. Rev. B* 8: 3719–3740

[14] Coehoorn R, Haas C, Dijkstra J, Flipse, De Groot RA and Wold A 1987 Electronic structure of MoSe$_2$, MoS$_2$, and WSe$_2$. I. Band-structure calculations and photoelectron spectroscopy. *Phys. Rev. B* 35: 6195–6202

[15] Ding Y, Wang Y, Ni J, Shi L, Shi S and Tang W 2011 First-principles study of structural, vibrational and electronic properties of graphene-like MX$_2$ (M=Mo, Nb, W, Ta; X=S, Se, Te) monolayers. *Physica B* 406: 2254–2260

[16] Manzeli S, Ovchinnikov D, Pasquier D, Yazyev OV and Kis A 2017 2D transition metal dichalcogenides. *Nature Rev. Mater.* 2: 1–15

[17] Wang QH, Kalantar-Zadeh K, Kis A, Coleman JN and Strano MS 2012 Electronics and optoelectronics of two-dimensional transition metal dichalcogenides. *Nature Nanotech.* 7: 699–712

[18] Coleman JN, Lotya M, O'Neill A, Bergin SD, King PJ, Khan U, *et al* 2011 Two-dimensional nanosheets produced by liquid exfoliation of layered materials. *Science* 331: 568-571

[19] Ayari A, Cobas E, Ogundadegbe O and Fuhrer MS 2007 Realization and electrical characterization of ultrathin crystals of layered transition-metal dichalcogenides. *J. Appl. Phys.* 101: 014507

[20] Mak KF, Lee C, Hone J, Shan J and Heinz TF 2010 Atomically thin $MoS_2$: A new direct-gap semiconductor. *Phys. Rev. Lett.* 105: 2–5

[21] Radisavljevic B, Radenovic A, Brivio J, Giacometti V and Kis A 2011 Single-layer $MoS_2$ transistors. *Nature Nanotech.* 6: 147–150

[22] Murray RB and Yoffe AD 1972 The band structures of some transition metal dichalcogenides: Band structures of the titanium dichalcogenides. *J. Phys. C* 5: 3038–3046

[23] Wei L, Jun-fang C, Qinyu H and Teng W 2010 Electronic and elastic properties of $MoS_2$. *Physica B* 405: 2498–2502

[24] Lebègue S and Eriksson O 2009 Electronic structure of two-dimensional crystals from ab initio theory. *Phys. Rev. B* 79: 4–7

[25] Ellis JK, Lucero MJ, and Scuseria GE 2011 The indirect to direct band gap transition in multilayered $MoS_2$ as predicted by screened hybrid density functional theory. *Appl. Phys. Lett.* 99: 261908

[26] Lee H, Zhang Q, Zhang W, Chang T, Lin T, Chang D, *et al* 2012 Synthesis of Large-Area $MoS_2$ Atomic Layers with Chemical Vapor Deposition. *Adv. Mater.* 24: 2320-2325

[27] Castellanos-Gomez A, Barkelid M, Goossens AM, Calado VE, van der Zant HSJ and Steele GA 2012 Laser-Thinning of $MoS_2$: On Demand Generation of a Single-Layer Semiconductor. *Nano Lett.* 12:187–3192

[28] Zhou SY, Gweon GH, Fedorov AV, First PN, de Heer WA, Lee DH, *et al* 2007 Substrate-induced bandgap opening in epitaxial graphene. *Nature Mater.* 6: 770-775

[29] Li X, Wang X, Zhang, Lee S and Dai H 2008 Chemically derived, ultrasmooth graphene nanoribbon semiconductors. *Science* 319: 1229–1232

[30] Han MY, Zyilmaz BO, Zhang Y and Kim P 2007 Energy Band-Gap Engineering of Graphene Nanoribbons. *Phys. Rev. Lett*. 98: 206805

[31] Jiao L, Zhang L, Wang X, Diankov G and Dai H 2009 Narrow graphene nanoribbons from carbon nanotubes. *Nature* 458: 877–880

[32] Singh N, Jabbour G and Schwingenschlögl U 2012 Optical and photocatalytic properties of two-dimensional $MoS_2$. *Eur. Phys. J. B* 85: 392

[33] Ataca C and Ciraci S 2011 Functionalization of single-layer $MoS_2$ honeycomb structures. *J. Phys. Chem. C* 115: 13303–13311

[34] Li H, Yin Z, He Q, Li H, Huang X, Lu G, *et al* 2012 Fabrication of single- and multilayer $MoS_2$ film-based field-effect transistors for sensing NO at room temperature. *Small* 8: 63-67

[35] Kim Y, Huang JL and Lieber CM 1991 Characterization of nanometer scale wear and oxidation of transition metal dichalcogenide lubricants by atomic force microscopy. *Appl. Phys. Lett.* 59: 3404–3406

[36] Hu KH, Hu XG and Sun XJ 2010 Morphological effect of $MoS_2$ nanoparticles on catalytic oxidation and vacuum lubrication. *Appl. Surf. Sci.* 256: 2517–2523

[37] Fortin E and Sears WM 1982 Photovoltaic effect and optical absorption in $MoS_2$. *J. Phys. Chem. Solids* 43: 881–884

[38] Wang PP, Sun H, Ji Y, Li W and Wang X 2014 Three-dimensional assembly of single-layered $MoS_2$. *Adv. Mater.* 26: 964–969

[39] Wu HC, Coileáin CÓ, Abid M, Mauit O, Syrlybekov A, Khalid A, *et al* 2015 Spin-dependent transport properties of $Fe_3O_4/MoS_2/Fe_3O_4$ junctions. *Sci Rep*. 5: 15984

[40] Dolui K, Narayan A, Rungger I and Sanvito S 2014 Efficient spin injection and giant magnetoresistance in Fe/MoS$_2$/Fe junctions. *Phys. Rev. B* 90: 1–5

[41] Zhang H, Ye M, Wang Y, Quhe R, Pan Y, Guo Y, *et al* 2016 Magnetoresistance in Co/2D MoS$_2$/Co and Ni/2D MoS$_2$/Ni junctions. *Phys. Chem. Chem. Phys.* 18: 16367-16376

[42] Rotjanapittayakul W, Pijitrojana W, Archer T, Sanvito S and Prasongkit J 2018 Spin injection and magnetoresistance in MoS$_2$-based tunnel junctions using Fe$_3$Si Heusler alloy electrodes. *Sci. Rep.* 8: 1–8

[43] Qian X, Liu J, Fu L and Li J 2014 Quantum spin hall effect in two - Dimensional transition metal dichalcogenides. *Science* 346: 1344–1347

[44] Zhu ZY, Cheng YC and Schwingenschlögl U 2011 Giant spin-orbit-induced spin splitting in two-dimensional transition-metal dichalcogenide semiconductors. *Phys. Rev. B* 84: 1–5

[45] Xiao D, Bin Liu G, Feng W, Xu X and Yao W 2012 Coupled spin and valley physics in monolayers of MoS$_2$ and other group-VI dichalcogenides. *Phys. Rev. Lett.*, 108: 1–5

[46] Ganatra R and Zhang Q 2014 Few-layer MoS$_2$: A promising layered semiconductor. *ACS Nano* 8: 4074–4099

[47] Lin MW, Kravchenko II, Fowlkes J, Li X, Puretzky AA, Rouleau CM, *et al* 2016 Thickness-dependent charge transport in few-layer MoS$_2$ field-effect transistors. *Nanotechnology* 27:165203

[48] Scalise E, Houssa M, Pourtois G, Afanas'ev V and Stesmans A 2012 Strain-induced semiconductor to metal transition in the two-dimensional honeycomb structure of MoS$_2$. *Nano Res.* 5: 43–48

[49] Dong L, Namburu RR, O'Regan TP, Dubey M and Dongare AM 2014 Theoretical study on strain-induced variations in electronic properties of monolayer MoS$_2$. *J. Mater. Sci.* 49:

6762–6771

[50] Zheng H, Yang B, Wang D, Han R, Du X and Yan Y 2014 Tuning magnetism of monolayer MoS$_2$ by doping vacancy and applying strain. *Appl. Phys. Lett.* 104: 1–6

[51] Tao P, Guo H, Yang T and Zhang Z 2014 Strain-induced magnetism in MoS$_2$ monolayer with defects. *J. Appl. Phys.* 115: 054305

[52] Chacko L, Swetha AK, Anjana R, Jayaraj MK and Aneesh PM, 2016 Wasp-waisted magnetism in hydrothermally grown MoS$_2$ nanoflakes. *Mater. Res. Express* 3: 1–9

[53] Zhang J, Soon JM, Loh KP, Yin J, Ding J, Sullivian MB *et al* 2007 Magnetic molybdenum disulfide nanosheet films. *Nano Lett.* 7: 2370–2376

[54] Andriotis AN and Menon M 2014 Tunable magnetic properties of transition metal doped MoS$_2$. *Phys. Rev. B* 90: 1–7

[55] Mishra R, Zhou W, Pennycook SJ, Pantelides ST and Idrobo JC 2013 Long-range ferromagnetic ordering in manganese-doped two-dimensional dichalcogenides. *Phys. Rev. B* 88: 1–5

[56] Ramasubramaniam A and Naveh D 2013 Mn-doped monolayer MoS$_2$: An atomically thin dilute magnetic semiconductor. *Phys. Rev. B* 87: 1–7

[57] Cheng YC, Zhu ZY, Mi WB, Guo ZB and Schwingenschlögl U 2013 Prediction of two-dimensional diluted magnetic semiconductors: Doped monolayer MoS$_2$ systems. *Phys. Rev. B* 87: 2–5

[58] Wang Y, Tseng L, Murmu PP, Bao N, Kennedy J, Ionesc M, *et al* 2017 Defects engineering induced room temperature ferromagnetism in transition metal doped MoS$_2$. *Mater. Design*, 121: 77-84

[59] Fang M and Yang E.-H. 2023 Advances in Two-Dimensional Magnetic Semiconductors via

Substitutional Doping of Transition Metal Dichalcogenides. *Materials* 16: 3701

[60] Lewis DJ, Tedstone AA, Zhong XL, Lewis EA, Rooney A, Savjani N, *et al* 2015 Thin Films of Molybdenum Disulfide Doped with Chromium by Aerosol-Assisted Chemical Vapor Deposition (AACVD). *Chem. Mater.* 27: 1367−1374

[61] Zhang R, Du Y, Han G and Gao X 2019 Ferromagnetism and microwave absorption properties of Cr-doped MoS$_2$ nanosheets. *J. Mater. Sci.* 54: 552–559

[62] He J, Wu K, Sa R, Li Q and Wei Y 2010 Magnetic properties of nonmetal atoms absorbed MoS$_2$ monolayers. *Appl. Phys. Lett.* 96: 82504

[63] Huang C, Jin Y, Wang W, Tang L, Song C and Xiu F 2017 Manganese and chromium doping in atomically thin MoS$_2$. *J. Semicond.* 38: 033004

[64] Kresse G and Furthmüller J 1996 Efficiency of ab-initio total energy calculations for metals and semiconductors using a plane-wave basis set. *Comput. Mater. Sci.* 6: 15–50

[65] Kresse G and Furthmüller J 1996 Efficient iterative schemes for ab-initio total-energy calculations using a plane-wave basis set. *Physical Review B* 54: 11169–11186

[66] Monkhorst HJ and Pack JD, 1976 Special points for Brillouin-zone integrations. *Phys. Rev. B* 13: 5188–5192

[67] Perdew JP and Zunger A 1981 Self-interaction correction to density-functional approximations for many-electron systems. *Phys. Rev. B* 23: 5048–5079

[68] Kresse G and Joubert D 1999 From Ultrasoft Pseudopotentials to the Projector Augmented-Wave Method. *Phys Rev B* 59: 1758–1775

[69] Perdew JP, Burke K and Ernzerhof M 1996 Generalized Gradient Approximation Made Simple, *Phys. Rev. Lett*. 77: 3865-3868

[70] Grimme S, Antony J, Ehrlich S and Krieg H 2010 A consistent and accurate ab initio

parametrization of density functional dispersion correction (DFT-D) for the 94 elements H-Pu. *J. Chem. Phys.* 132: 154104

[71] Dudarev SL, Botton GA, Savrasov SY, Humphreys CJ and Sutton AP, 1998 Electron-energy-loss spectra and the structural stability of nickel oxide: An LSDA+U study. *Phys. Rev. B* 57: 1505–1509

[72] Anisimov VI, Zaanen J and Andersen O K 1991 Band theory and Mott insulators: Hubbard *U* instead of Stoner *I*. *Phys. Rev. B* 44: 943–954

[73] Krukau AV, Vydrov OA, Izmaylov AF and Scuseria GE 2006 Influence of the exchange screening parameter on the performance of screened hybrid functionals. *J. Chem. Phys.* 125: 224106

[74] Wu M, Yao X, Hao Y, Dong H, Cheng Y, Liu H, *et al* 2018 Electronic structures, magnetic properties and band alignments of 3d transition metal atoms doped monolayer $MoS_2$. *Phys. Lett. A* 382: 111-115

[75] Conley HJ, Wang B, Ziegler JI, Haglund RF, Pantelides ST and Bolotin KI 2013 Bandgap Engineering of Strained Monolayer and Bilayer $MoS_2$. *Nano Lett.* 13: 3626-3630

**Figure and Table Captions**

**Figure 1**. The (a) band structure and (b) atom projected density of states of bulk 2H-MoS$_2$.

**Figure 2**. Band structure of bulk MoS$_2$ with (a) DFT and (b) HSE method.

**Figure 3**. Atom projected DOS for undoped MoS$_2$ (a) monolayer, (b) bilayer, (c) trilayer, (d) four-layer.

**Figure 4**. Band structure of mono-layer (a), bilayer (b), trilayer (c), and four-layer (d) MoS$_2$.

**Figure 5**. The unique possible positions for two Cr dopants for (a) in-plane (b) out-of-plane configurations in a 4×4×1 supercell of MoS$_2$. The in-plane case has three configurations, whereas the out-plane case has four unique configurations. The Mo atoms are in violet and S in yellow, while other colors are described in the text.

**Figure 6**. The electronic density of states (a) and band structure for Cr-doped bulk MoS$_2$ (b).

**Figure 7**. The band structure for a monolayer of MoS$_2$ (a) pristine and (b) two Cr atoms substituted at Mo sites. In the doped case, the contribution due to Cr is shown by bands with red colour and green for Mo atoms.

**Figure 8**. The DOS of the Cr-doped in-plane (a) bilayer and (b) trilayer and (c) four-layer MoS$_2$ and out-of-plane MoS$_2$ (d) bilayer (e) trilayer and (f) four-layer configurations, respectively.

**Figure 9**. Band structure of Cr-doped in the in-plane of MoS$_2$ monolayer and multilayers, (a) monolayer, (b) bilayer, (c) trilayer, (d) four-layer.

**Figure 10**. Band structure of Cr-doped out-of-plane configurations of MoS$_2$ multilayers, (a) for bilayer, (b) trilayer, and (c) four-layer.

**Figure 11**. Atom projected band structure of in-plane Cr-doped MoS$_2$ monolayers (a) and (b), and out-of-plane Cr-doped bilayer MoS$_2$ (c) and (d). Green lines denote energy levels for Mo, and (b) red lines for energy levels for the Cr atom.

**Figure 12**. Band structure of monolayer $MoS_2$ with (a) DFT and (b) HSE06 hybrid functional calculation. With HSE06 hybrid functional, the conduction band shifts by 0.25 eV, and the energy levels are correctly described, which can be seen with the splitting of conduction band energy levels. The VBM and CBM are at the K-point, so the band gap is direct.

**Figure 13**. Band structure of Cr-doped monolayer $MoS_2$ with (a) DFT and (b) HSE06 calculation. With HSE06 hybrid functional, the Cr-defect level shifts upward and forms around 1.5 eV. The VBM is at the K point, whereas the CBM is at the K point in (a) and in between the K and $\Gamma$ points in (b).

**Table 1**. The energy band gap of various 2D and bulk $MoS_2$. The monolayer has a direct band gap, whereas it is indirect for the bulk and multilayers.

**Table 2**. The total energy for various configurations in the in-plane and out-of-plane doping.

**Table 3**. Magnetic moments and band gaps obtained for different U parameter values in DFT+U calculations for Cr-doped $MoS_2$ monolayer with the Cr $3p$-orbitals in the core.